\documentclass[letterpaper]{article}
\usepackage{helvet}
\usepackage{courier}
\usepackage[latin9]{inputenc}
\usepackage{mathtools}
\usepackage{multirow}
\usepackage{amsmath}
\usepackage{graphicx}
\usepackage[authoryear]{natbib}

\makeatletter

\providecommand{\tabularnewline}{\\}

\def\year{2021}\relax
\usepackage{aaai21}% DO NOT CHANGE THIS
\usepackage{times}% DO NOT CHANGE THIS
\usepackage[hyphens]{url}% DO NOT CHANGE THIS
\usepackage{amsfonts}% blackboard math symbols

\usepackage{caption}
\usepackage{caption}
\title{Analyzing Epistemic and Aleatoric Uncertainty for Drusen Segmentation in Optical Coherence Tomography Images}
\author{ Tinu Theckel Joy, Suman Sedai, Rahil Garnavi
}
\affiliations{
    
 IBM Research - Australia, Melbourne, VIC, Australia \\
}
\@ifundefined{showcaptionsetup}{}{%
 \PassOptionsToPackage{caption=false}{subfig}}
\usepackage{subfig}
\makeatother

\begin{document}
\maketitle 
\begin{abstract}
Age-related macular degeneration (AMD) is one of the leading causes
of permanent vision loss in people aged over 60 years. Accurate segmentation
of biomarkers such as drusen that points to the early stages of AMD
is crucial in preventing further vision impairment. However, segmenting
drusen is extremely challenging due to their varied sizes and appearances,
low contrast and noise resemblance. Most existing literature, therefore,
have focused on size estimation of drusen using classification, leaving
the challenge of accurate segmentation less tackled. Additionally,
obtaining the pixel-wise annotations is extremely costly and such
labels can often be noisy, suffering from inter-observer and intra-observer
variability. Quantification of uncertainty associated with segmentation
tasks offers principled measures to inspect the segmentation output.
Realizing its utility in identifying erroneous segmentation and the
potential applications in clinical decision making, here we develop
a U-Net based drusen segmentation model and quantify the segmentation
uncertainty. We investigate epistemic and aleatoric uncertainty capturing model
confidence and data uncertainty respectively. We present segmentation results and show how uncertainty can help
formulate robust evaluation strategies. We visually inspect the pixel-wise
uncertainty and segmentation results on test images. We finally analyze
the correlation between segmentation uncertainty and accuracy. Our
results demonstrate the utility of leveraging uncertainties in developing
and explaining segmentation models for medical image analysis. 
\end{abstract}

\section{Introduction\label{sec:intro}}

Age-related macular degeneration (AMD) is a retinal disease leading
to permanent vision loss among the aged population worldwide. Early
detection of AMD plays a vital role in limiting the disease progression.
Drusen - protein or lipid deposits that usually accumulate between
Bruch's membrane and retinal epithelium layer - is a key biomarker
that can indicate early developments of AMD. There are efforts aimed
at detecting drusen in fundus and optical coherence tomography (OCT)
images \cite{peng2019deepseenet,saha2019automated}, however, most
have focused mainly on the volumetric classification of drusen, categorizing
it into different size \cite{keenan2020deep}. There are also some
effort in classification of AMD based on absence or presence of any
pathology \cite{antony2019automated}. A challenging and rarely attempted
task is the precise segmentation a.k.a pixel-wise classification of
the drusen in OCT images. This segmentation task is particularly challenging
due to the drusen appearing in varied sizes and shapes, its resemblance
to noise which is commonly present in OCT, as well as the low contrast.

Segmentation is ubiquitous in many areas of medical imaging. Over
the years, convolutional neural networks (CNN) and its variations
like U-Net \cite{ronneberger2015u,li2018h} have shown promising
results on many medical images. However, deep learning algorithms
applied to different segmentation tasks in medical imaging have not
enjoyed the same level of success as natural images. The primary reason
for such an impediment is the difficulty of having access to a large
amount of high-quality labelled data. Annotating medical images is
costly as it often requires the expertise of clinicians. In ambiguous
cases, even expert annotations can be inconsistent, resulting in noisy
annotations. The model will produce inaccurate segmentation when trained
using such limited and noisy data. Quantification of uncertainties
associated with the segmentation output is therefore important to
determine the region of possible incorrect segmentation, e.g., region
associated with higher uncertainty can either be excluded from subsequent
analysis or highlighted for manual attention.

Epistemic and aleatoric uncertainty are two major types of uncertainty
that one can quantify in Bayesian deep learning \cite{der2009aleatory,kendall2017uncertainties}.
Epistemic uncertainty captures the uncertainty in the model parameters
due to the lack of knowledge about the underlying model that generated
the given data. In contrast, aleatoric uncertainty captures noise
inherent in the input data. While the epistemic uncertainty can be
reduced by collecting more training data, aleatoric uncertainty can
not be explained away by having more data \cite{kendall2017uncertainties}.

The literature on uncertainty quantification in the context of diagnosing
AMD mainly aim at analyzing the epistemic uncertainty for the relatively
easier task of segmenting the retinal layers in OCT images \cite{sedai2019uncertainty,seebock2019exploiting}.
We believe no work in the literature has attempted the task of quantifying
and analyzing both the epistemic and aleatoric uncertainty associated
with segmenting the drusen in OCT images. Hence, there exists a gap
in the literature on developing a segmentation model for detecting
drusen and quantifying the uncertainty towards building a robust system
for early diagnosis of AMD.

In this quest, we develop a U-Net \cite{li2018h} based segmentation
framework for detecting the drusen in OCT images on a benchmark dataset.
We model the segmentation uncertainty using both epistemic and aleatoric
uncertainty measures. We evaluate the generalization performance of
the model for drusen segmentation. We further show the utility of
segmentation uncertainty in evaluating the model on specific regions
of test images. We visualize the drusen segmentation and pixel-wise
uncertainty measures on test images. We conclude by analyzing the
association between segmentation uncertainty and accuracy. Our results
demonstrate that both epistemic and aleatoric uncertainty helps to
explain the erroneous region of drusen segmentation at test-time.

\section{Drusen Segmentation and Uncertainty Quantification\label{sec:framework} }

%\vspace{-0.5cm}
Having only a limited number of labelled data - which is often the
reality of developing deep learning models in medical applications - we augment
the training set by introducing different patches containing drusen
from a single image. We create the patches by cropping an image with
windows of sizes 128, 192, and 256. This technique of lowering the
resolution of the image also helps us to avoid resizing the images
to preserve important aspects of the pathologies. To be consistent
with the training data, we also generate patches from test data by
using a window of size 128.

We use a standard U-Net \cite{li2018h} model with the encoder-decoder
architecture. Both the encoder and decoder have four blocks wherein
each block in the encoder comprises of four convolutional units followed
by batch normalization \cite{ioffe2015batch} and leaky rectified
linear unit \cite{maas2013rectifier}. We use skip connections \cite{drozdzal2016importance}
between the output of the encoder blocks and input of the decoder
blocks. In the final layer, we use a convolution layer with channel
number equal to the number of classes and a softmax activation function.
In our U-Net architecture, we use spatial dropout \cite{tompson2015efficient}
before every convolutional layer. We train the model and subsequently
compute the segmentation and uncertainty maps.

We compute the pixel-wise epistemic uncertainty using Monte-Carlo
dropout \cite{gal2016dropout} which characterizes the dropout regularization
(spatial dropout in our model) as a variational Bayesian inference
problem. The Monte-Carlo dropout \cite{gal2016dropout} quantifies
the uncertainty by having $T$ stochastic forward passes during inference
where the dropout is enabled at each pass. For a given class $c$
and pixel $x$ in an input image, we obtain the output by averaging
the softmax probabilities over multiple forward passes as, 
\begin{equation}
p(y=c\mid x,D)=\dfrac{1}{T}\sum_{t=1}^{^{T}}p(y=c\mid x,\boldsymbol{w}_{t})\label{eq:epistemic}
\end{equation}
where $t=1,2,...,T$ denotes each forward pass, $\boldsymbol{w}_{t}$
denotes the weights of the model after applying dropout at $t^{th}$
pass and $D$ is the training dataset. We use $T=10$ in our experiments.

Quantifying the variations in the predictions by augmenting data at
test-time is a simple yet effective strategy for estimating aleatoric
uncertainty \cite{ayhan2018}. Motivated by this, we quantify the
pixel-wise aleatoric uncertainty for drusen segmentation using test-time
augmentation. During inference, we feed the model with $T$ transformations
of the input image generated using different augmentation techniques
to obtain probability distribution over the predictions. Similar to
epistemic, we average the softmax probabilities over multiple transformations
as, 
\begin{equation}
p(y=c\mid x,D)=\dfrac{1}{T}\sum_{t=1}^{^{T}}p\left(y=c\mid M_{t}^{-1}(M_{t}(x)),\boldsymbol{w}\right)\label{eq:aleatoric}
\end{equation}
where $M_{t}$ denotes the transformation operation at $t$ and $\boldsymbol{w}$
is the model weight. The reverse transformation $M_{t}^{-1}$ is only
applied in case of geometrical augmentation techniques like rotation.
Additionally, $M_{t}^{-1}$ is an identity operation for image processing
based transformations like blurring. The current work uses randomized
augmentation techniques that include adjusting brightness, contrast,
blurring and rotation of the images.

We evaluate the uncertainty for a pixel $x$ using the entropy of
predictive probability distribution as, 
\begin{equation}
\mathbb{H}\left[y\mid x,D\right]\coloneqq-\sum_{c}p(y=c\mid x,D)\log(p(y=c\mid x,D))
\end{equation}
where $p(y)$ is the average of softmax probabilities computed using
Equations (\ref{eq:epistemic}) and (\ref{eq:aleatoric}) for epistemic
and aleatoric methods respectively. Similarly, we compute the average
segmentation uncertainty of drusen as, 
\begin{equation}
U_{avg}=\dfrac{1}{N}\sum_{p(y)\geq0.5}\mathbb{H}\left[y\mid x,D\right]
\end{equation}
where $N$ is the number of pixels that satisfy $p(y)\geq0.5$.
\begin{table*}[p]
\noindent \centering{}%
\begin{tabular}{|c|c|c|c|c|c|c|c|c|c|}
\hline 
\multirow{2}{*}{Methods} & \multicolumn{3}{c|}{Large} & \multicolumn{3}{c|}{Medium} & \multicolumn{3}{c|}{Small}\tabularnewline
\cline{2-10} \cline{3-10} \cline{4-10} \cline{5-10} \cline{6-10} \cline{7-10} \cline{8-10} \cline{9-10} \cline{10-10} 
 & Dice & Precision  & Recall  & Dice & Precision  & Recall  & Dice  & Precision  & Recall\tabularnewline
\hline 
{{}{}{}{}{}{}}no-uncertainty{{} }  & {{}{}{}{}{}{}0.72 }  & {{}0.83}  & {{}0.67}  & {{}{}{}{}{}{}0.65 }  & {{}0.71}  & {{}0.61}  & {{}{}{}{}{}{}0.55}  & {{}0.63}  & {{}0.51}\tabularnewline
\hline 
{{}{}{}{}{}{}}epistemic{{} }  & {{}{}{}{}{}{}0.72 }  & {{}0.84}  & {{}0.65}  & {{}{}{}{}{}{}0.64 }  & {{}0.72}  & {{}0.58}  & {{}{}{}{}{}{}0.53}  & {{}0.68}  & {{}0.48}\tabularnewline
\hline 
{{}{}{}{}{}{}}aleatoric{{} }  & {{}{}{}{}{}{}0.73 }  & {{}0.83}  & {{}0.67}  & {{}{}{}{}{}{}0.64 }  & {{}0.71}  & {{}0.6}  & {{}{}{}{}{}{}0.54}  & {{}0.64}  & {{}0.5}\tabularnewline
\hline 
{{}{}{}{}{}{}}epistemic-thresholded{{} }  & {{}{}{}{}{}{}0.8}  & {{}0.93}  & {{}0.72}  & {{}{}{}{}{}{}0.71 }  & {{}0.85}  & {{}0.64}  & {{}{}{}{}{}{}0.57}  & {{}0.75}  & {{}0.5}\tabularnewline
\hline 
{{}{}{}{}{}{}}aleatoric-thresholded{{} }  & {{}{}{}{}{}{}0.8}  & 0.91  & 0.74  & {{}{}{}{}{}{}0.71}  & 0.82  & 0.63  & {{}{}{}{}{}{}0.57}  & 0.72  & 0.5\tabularnewline
\hline 
\end{tabular}\caption{Model performance on holdout test dataset: average of the scores from
different methods are reported across large, medium and small-sized
drusen. The method "no-uncertainty" indicates U-Net model which
do not estimate any uncertainty. In the last two methods, the pixels
having higher uncertainty (least confidence) are excluded from the
evaluation.\label{tab:Model-performance-on}}
\end{table*}

\section{Results\label{sec:exp}}

We conduct experiments on a benchmark dataset that is publicly available
\cite{Farsu2014}. The dataset consists of OCT volumes for 269 subjects
diagnosed with AMD and 115 normal subjects. We extract 286 images
from the 143 AMD OCT volumes by sampling B-scans which have visible
drusen. We then obtain the expert annotations for the drusen in all
the images. We create a training set with images sampled from 70\%
of the subjects and use the remaining for validation. We augment the
training data by flipping the images horizontally and rotating them.
We evaluate the model on a holdout test dataset. To gain more insights
into the generalization performance of the model across varied sizes
of drusen, we divide the images into large, medium, and small based
on the size of drusen. 
\begin{figure*}[p]
\begin{centering}
\subfloat[Input]{\begin{centering}
\includegraphics[scale=0.36]{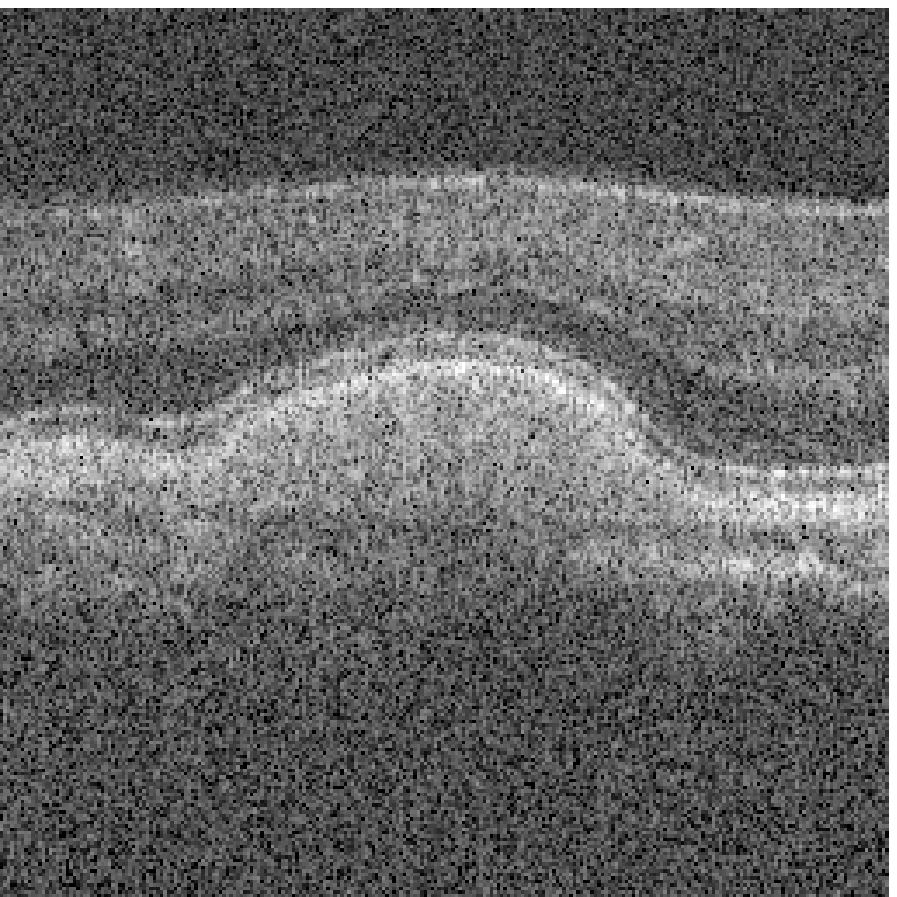} 
\par\end{centering}
}\subfloat[Ground Truth]{\begin{centering}
\includegraphics[scale=0.36]{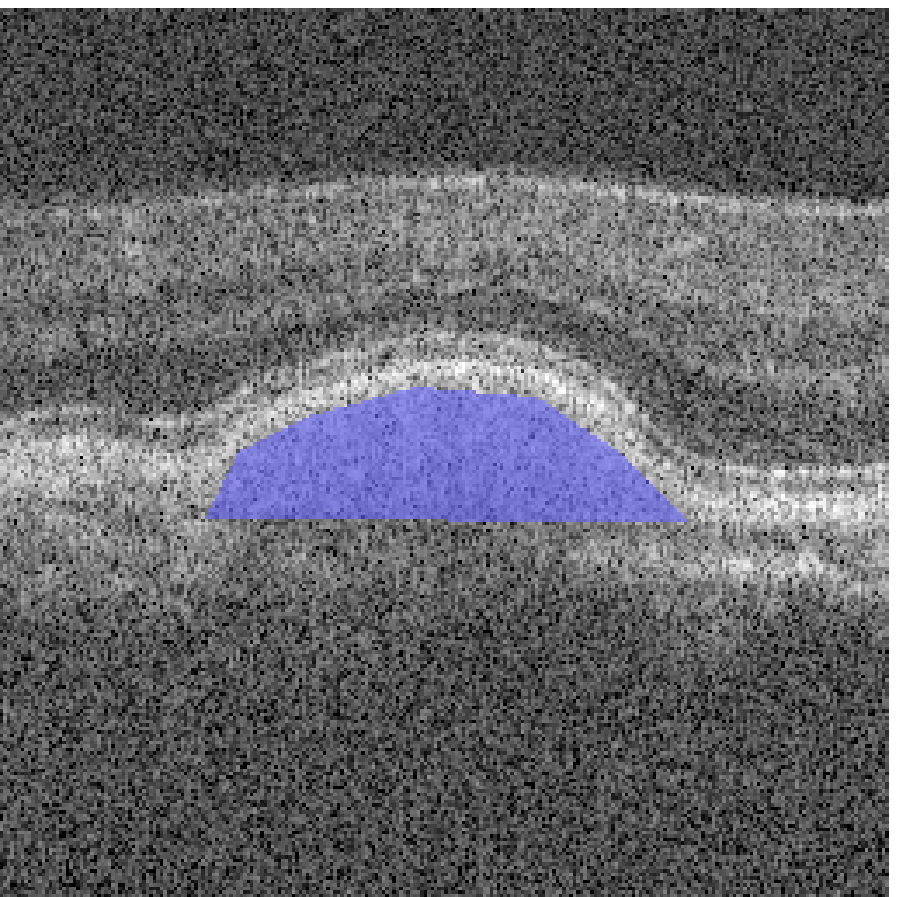} 
\par\end{centering}
}\subfloat[Segmentation]{\begin{centering}
\includegraphics[scale=0.36]{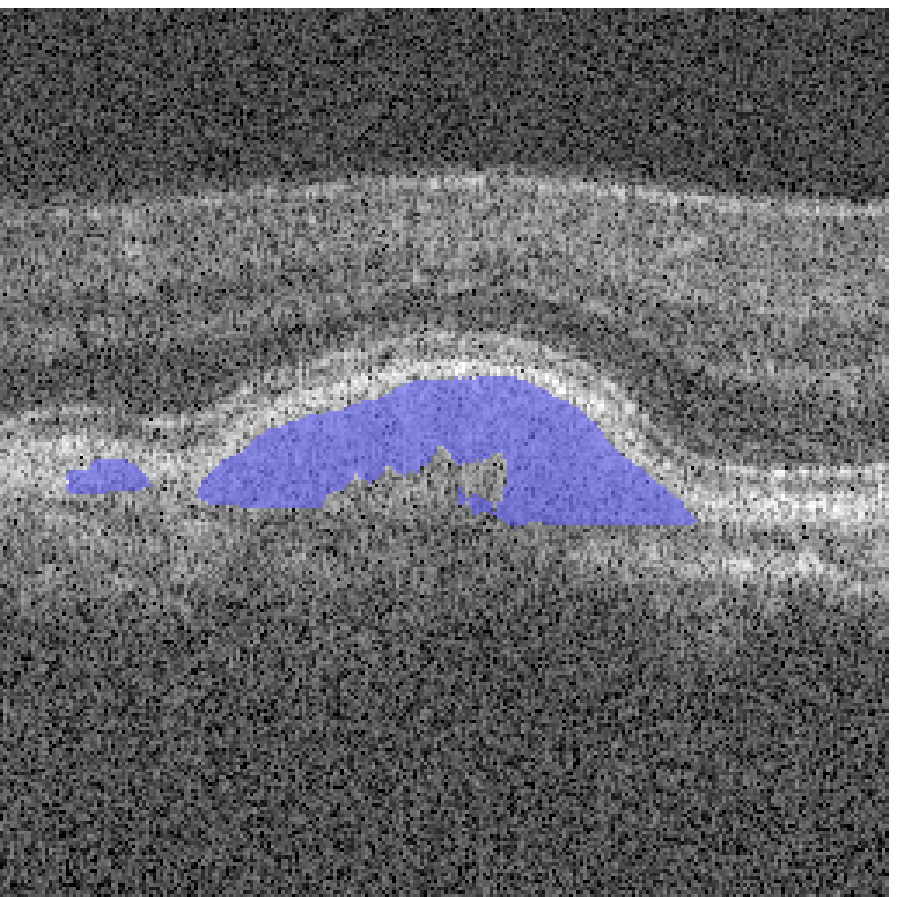} 
\par\end{centering}
}\subfloat[Epistemic Uncertainty]{\begin{centering}
\includegraphics[scale=0.34]{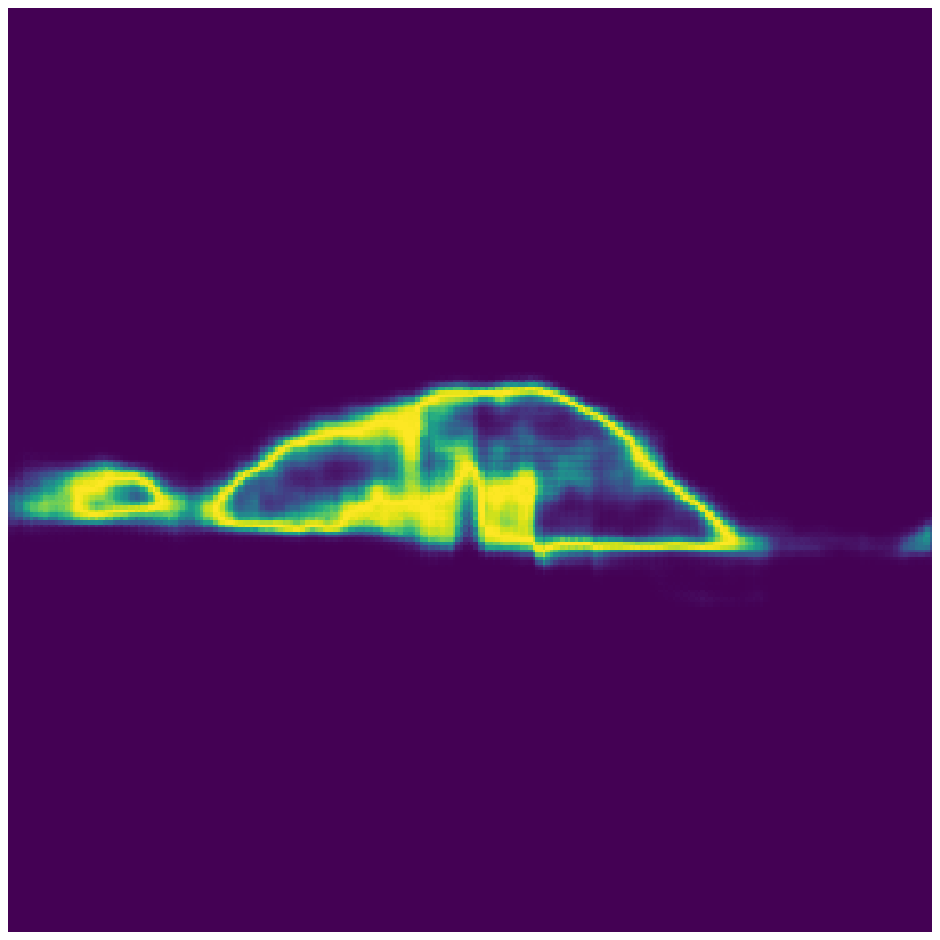} 
\par\end{centering}
}\subfloat[Aleatoric Uncertainty]{\begin{centering}
\includegraphics[scale=0.34]{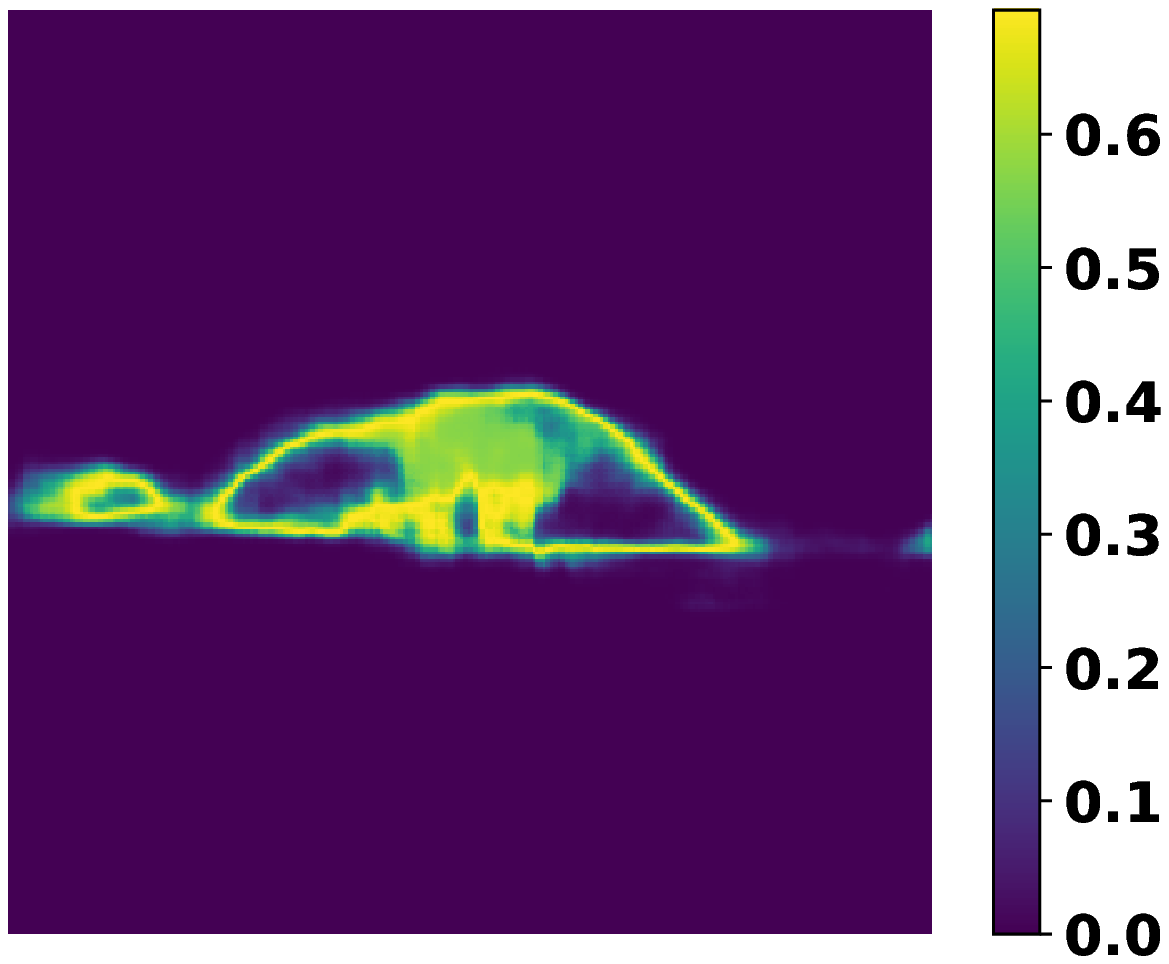} 
\par\end{centering}
}
\par\end{centering}
\begin{centering}
\subfloat[]{\begin{centering}
\includegraphics[scale=0.36]{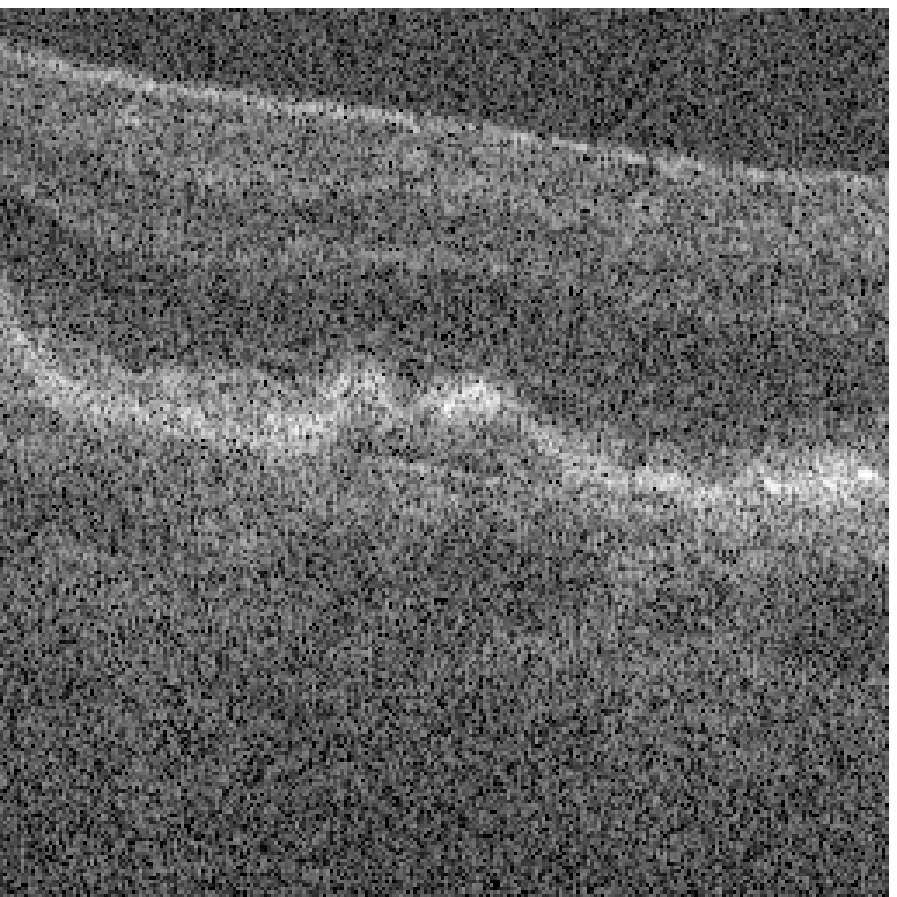} 
\par\end{centering}
}\subfloat[]{\begin{centering}
\includegraphics[scale=0.36]{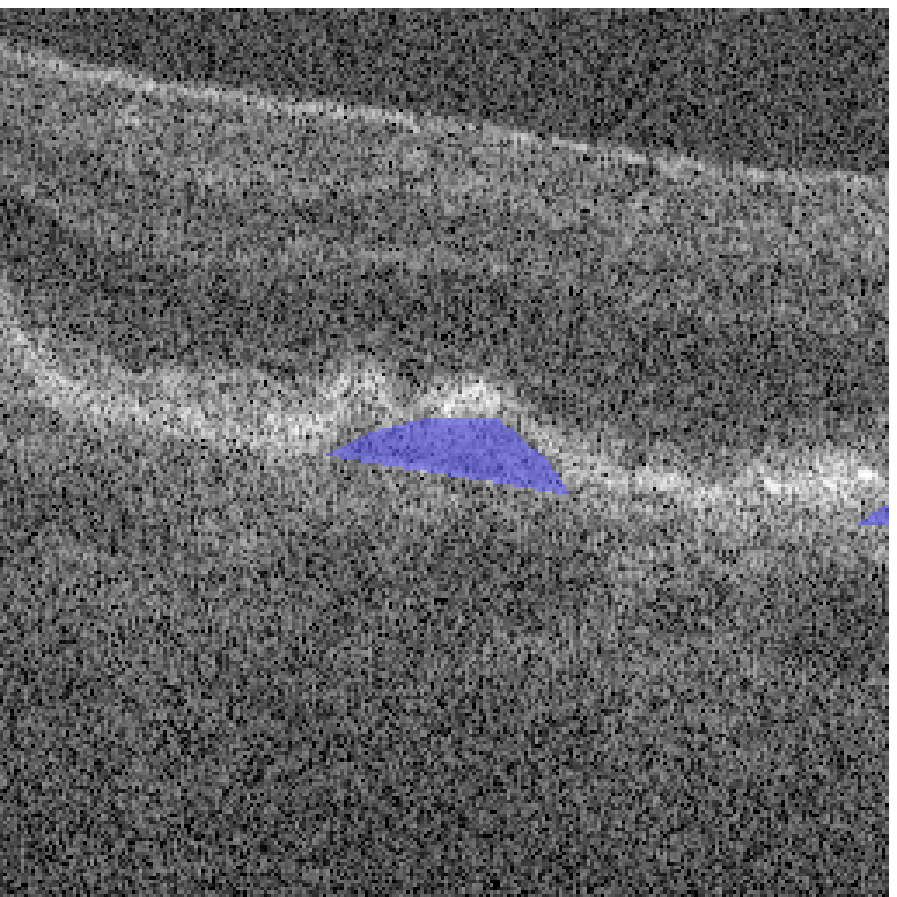} 
\par\end{centering}
}\subfloat[]{\begin{centering}
\includegraphics[scale=0.36]{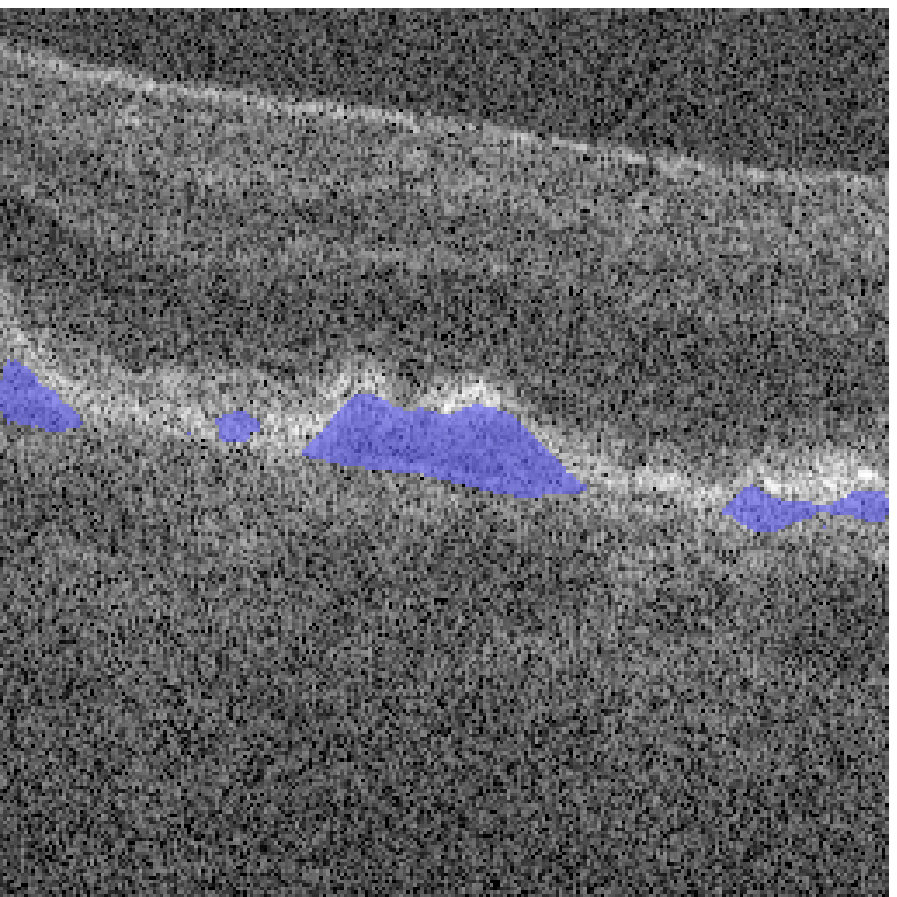} 
\par\end{centering}
}\subfloat[]{\begin{centering}
\includegraphics[scale=0.34]{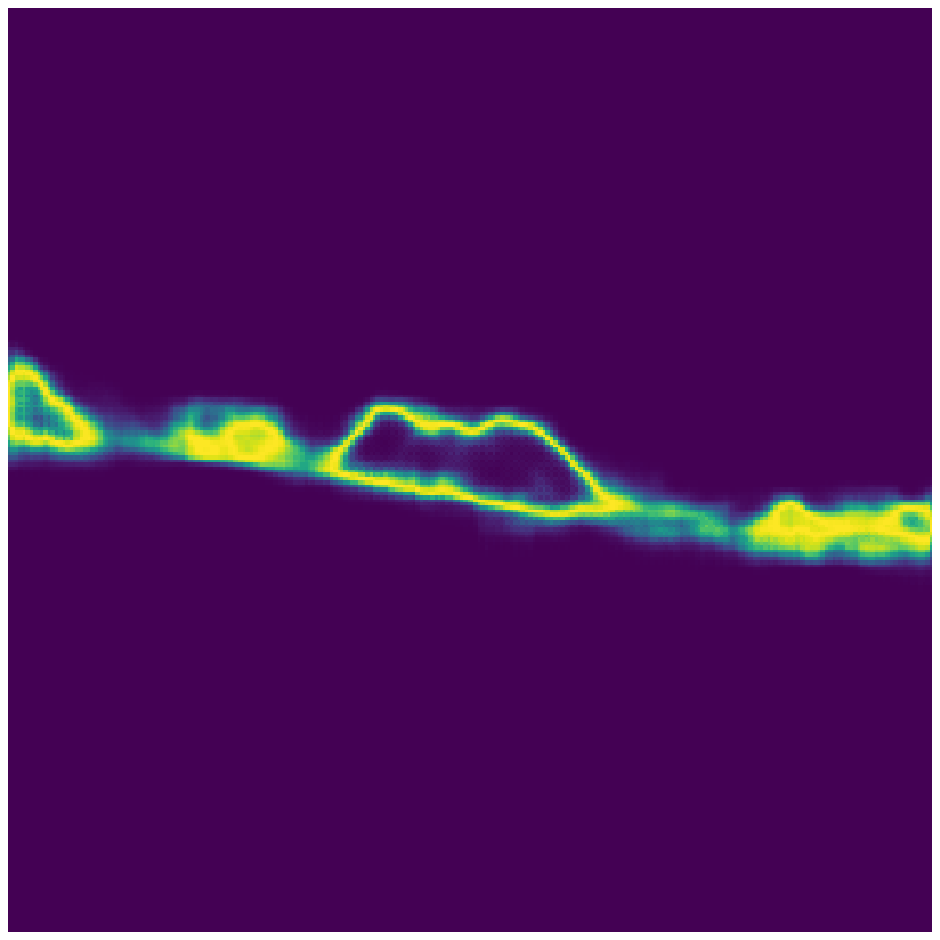} 
\par\end{centering}
}\subfloat[]{\begin{centering}
\includegraphics[scale=0.34]{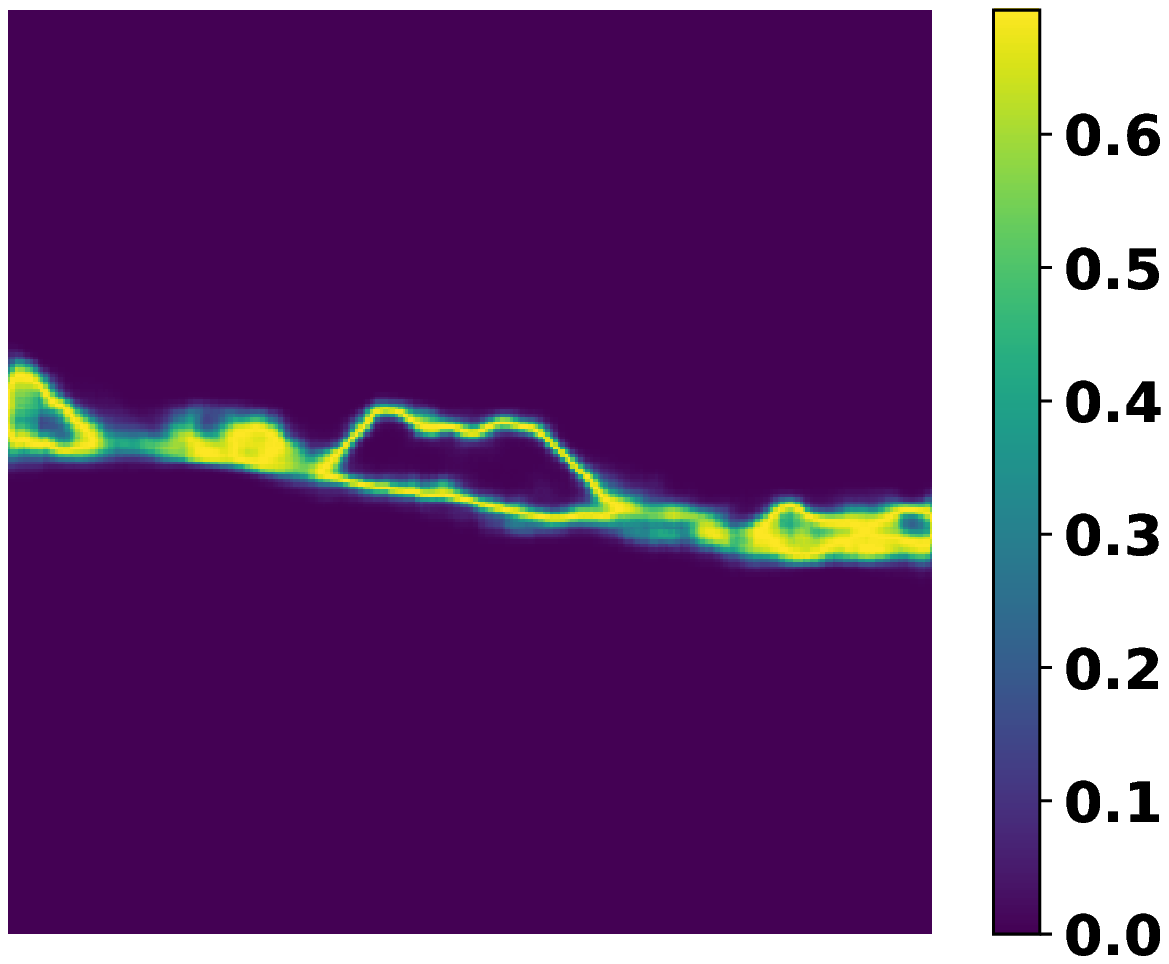} 
\par\end{centering}
}
\par\end{centering}
\begin{centering}
\subfloat[]{\begin{centering}
\includegraphics[scale=0.36]{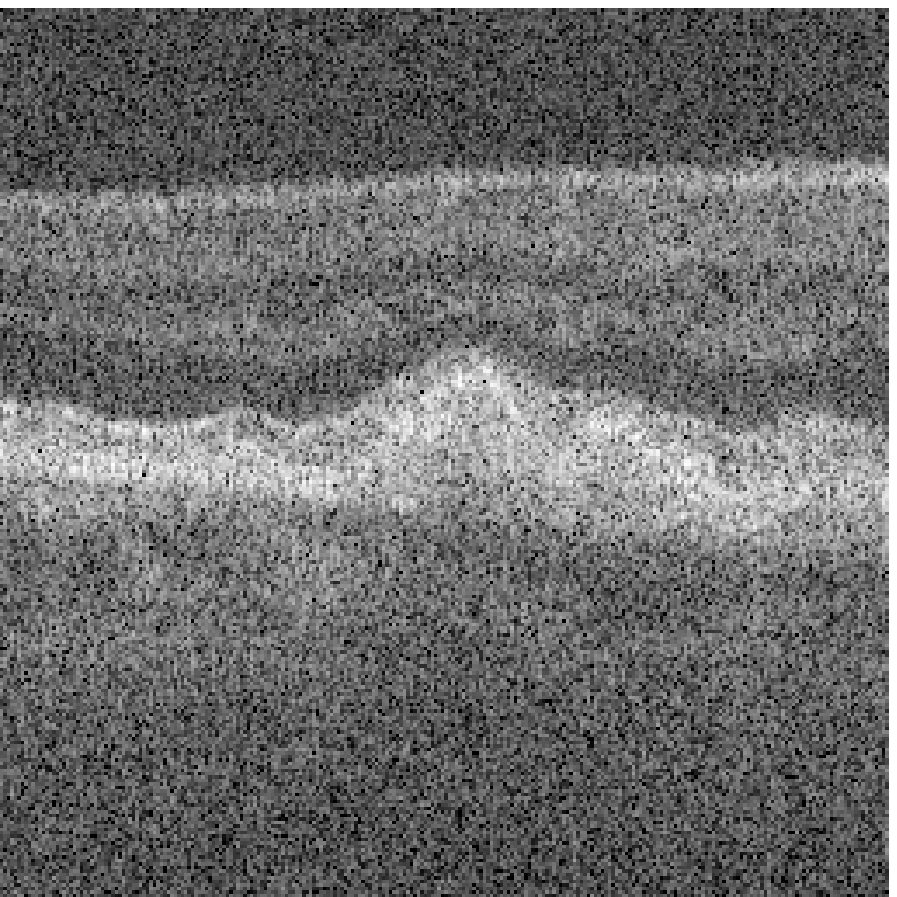} 
\par\end{centering}
}\subfloat[]{\begin{centering}
\includegraphics[scale=0.36]{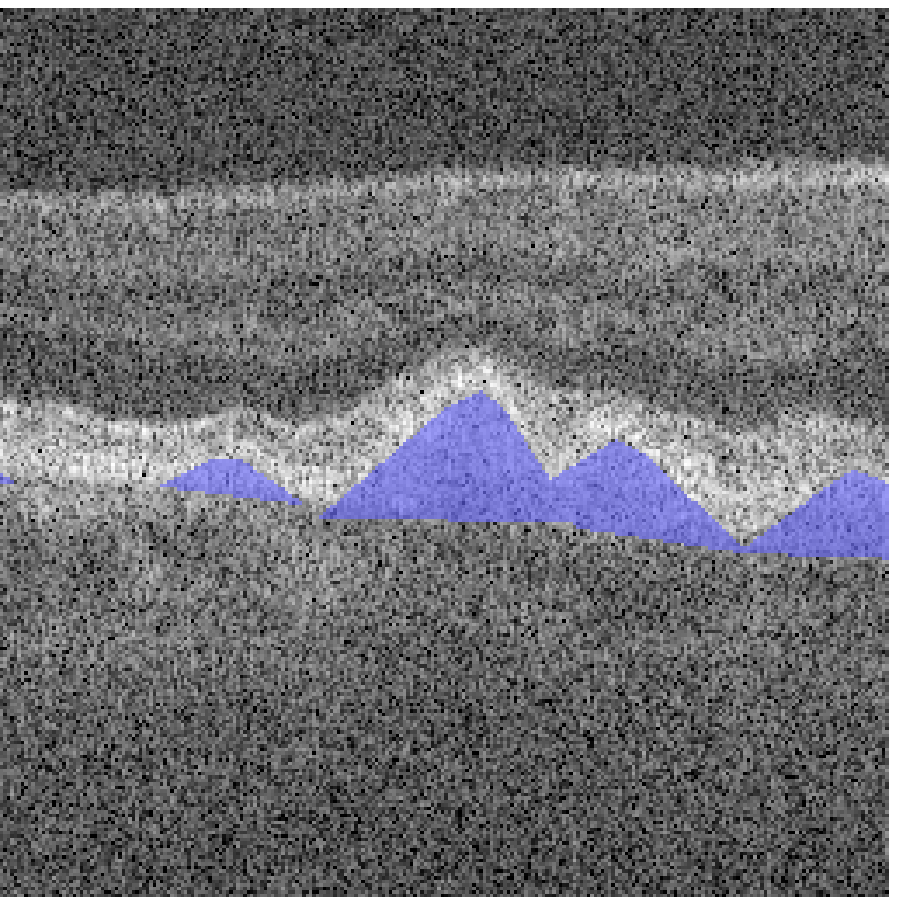} 
\par\end{centering}
}\subfloat[]{\begin{centering}
\includegraphics[scale=0.36]{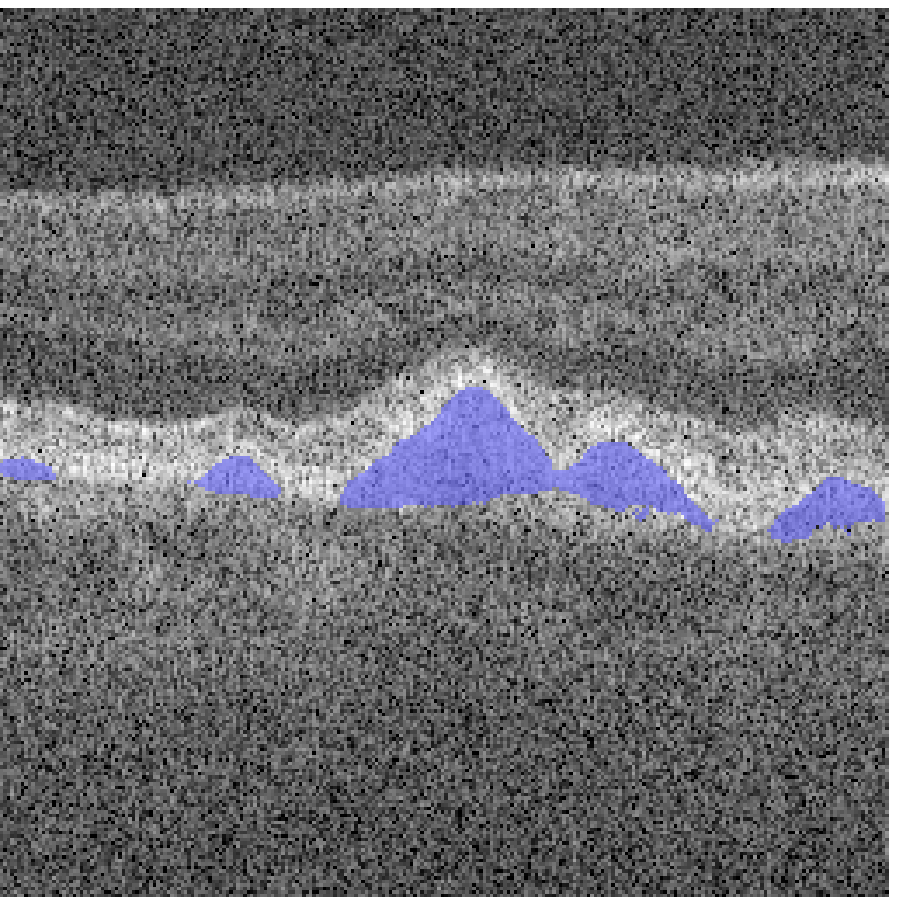} 
\par\end{centering}
}\subfloat[]{\begin{centering}
\includegraphics[scale=0.34]{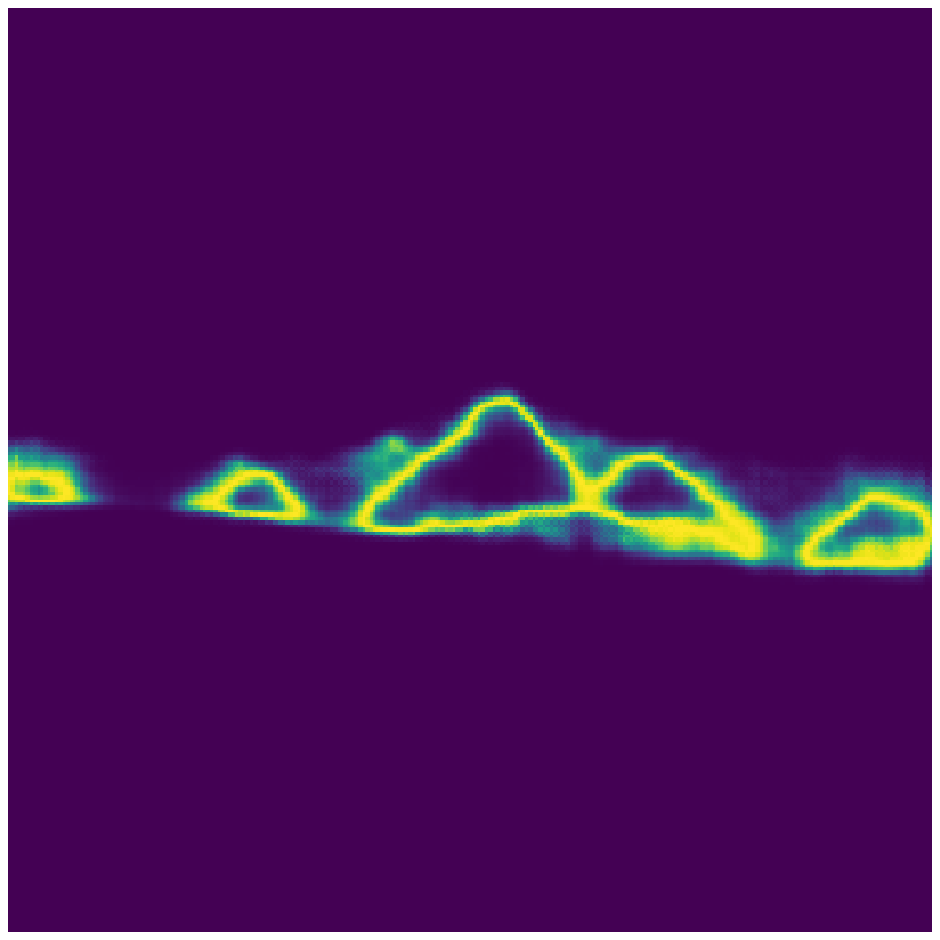} 
\par\end{centering}
}\subfloat[]{\begin{centering}
\includegraphics[scale=0.34]{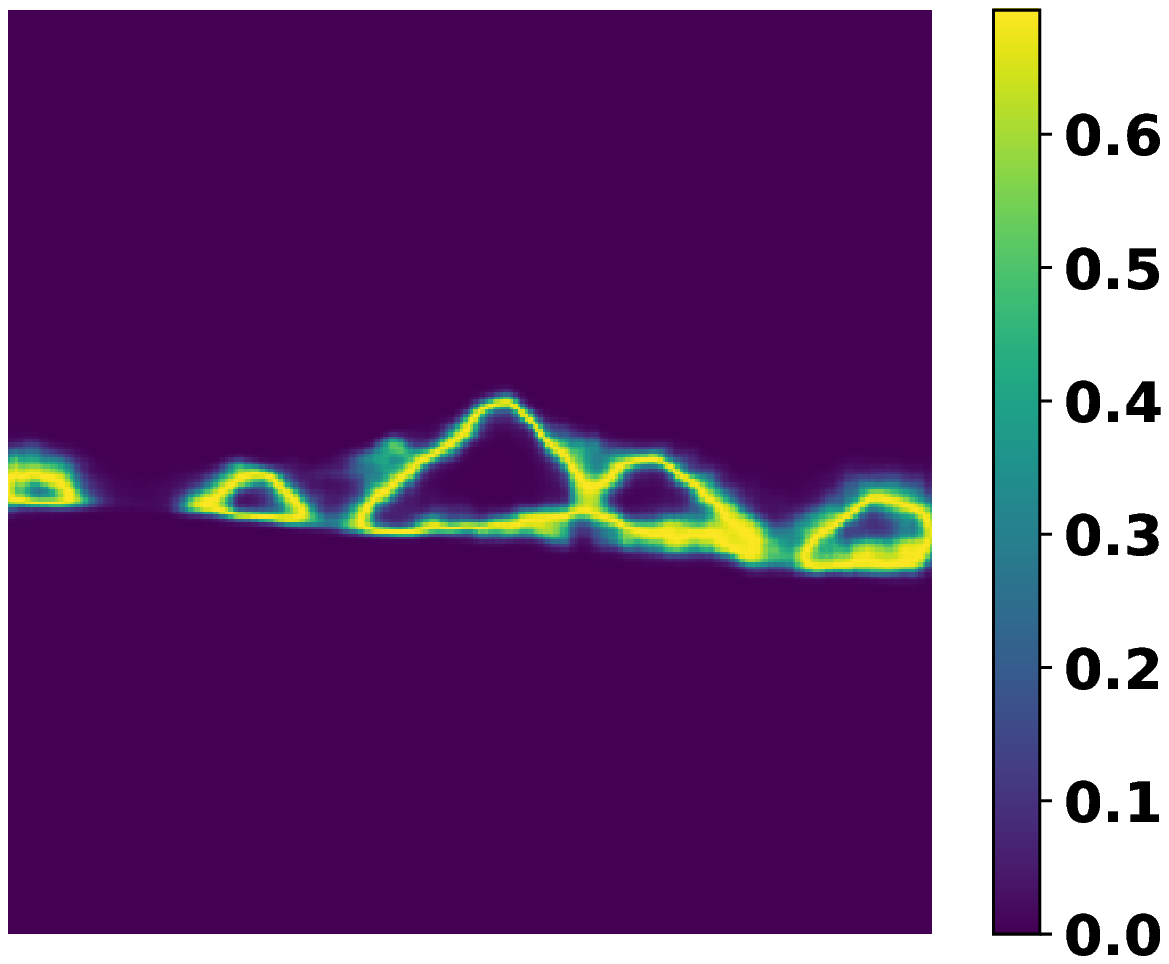} 
\par\end{centering}
}
\par\end{centering}
\begin{centering}
\subfloat[]{\begin{centering}
\includegraphics[scale=0.36]{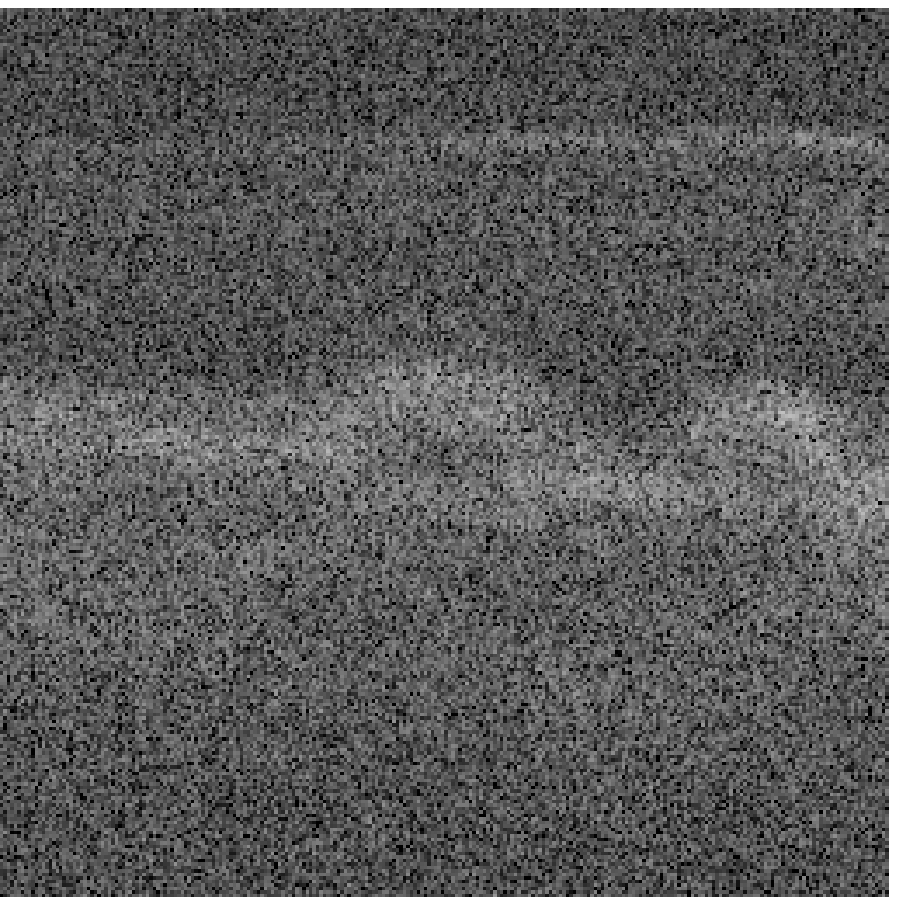} 
\par\end{centering}
}\subfloat[]{\begin{centering}
\includegraphics[scale=0.36]{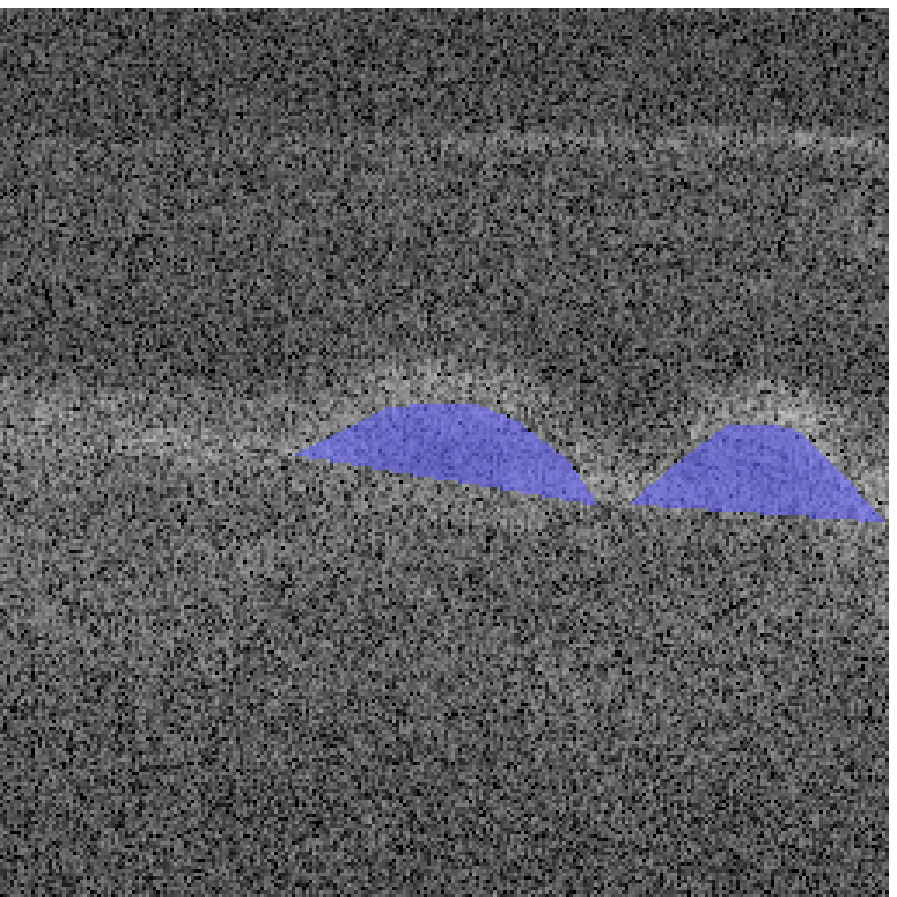} 
\par\end{centering}
}\subfloat[]{\begin{centering}
\includegraphics[scale=0.36]{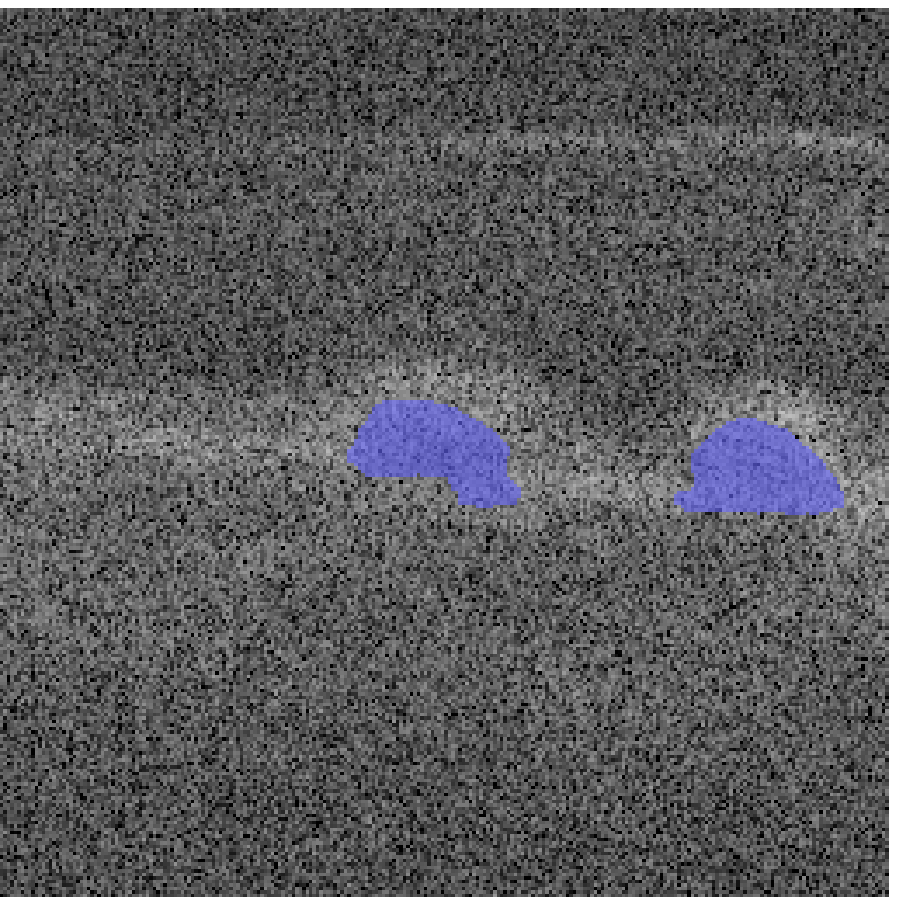} 
\par\end{centering}
}\subfloat[]{\begin{centering}
\includegraphics[scale=0.34]{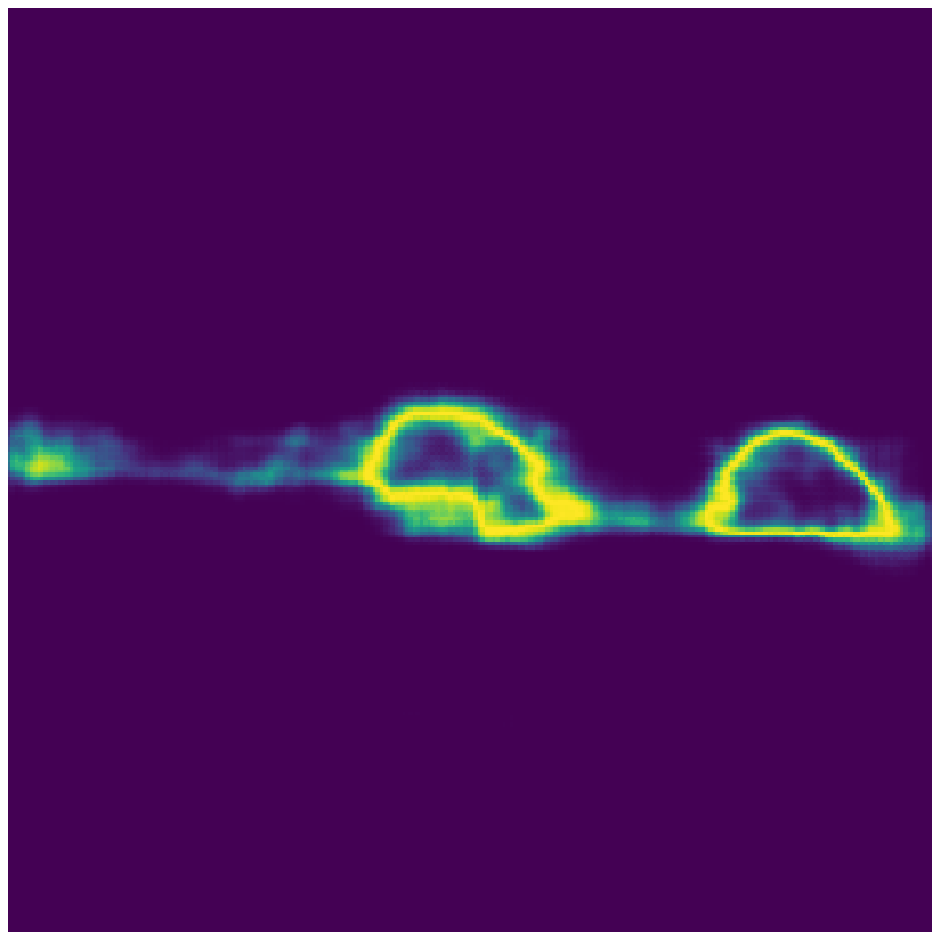} 
\par\end{centering}
}\subfloat[]{\begin{centering}
\includegraphics[scale=0.34]{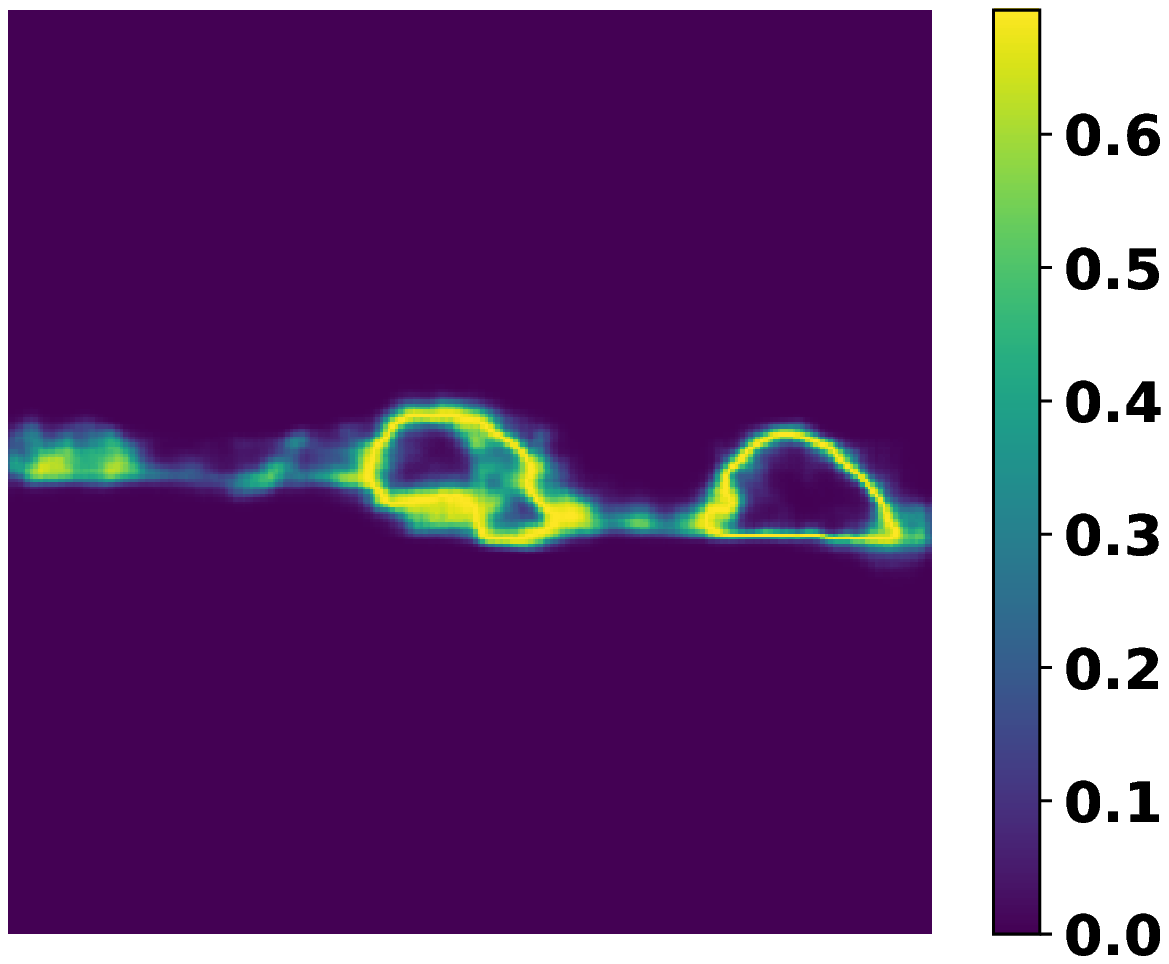} 
\par\end{centering}
}
\par\end{centering}
\caption{Visualization of drusen segmentation with pixel-wise epistemic and
aleatoric uncertainty on test images. Ground truth and predicted drusen
pixels are highlighted with a separate colour.\label{fig:Visualization-of-epistemic}}
\end{figure*}

\begin{figure*}[t]
\begin{centering}
\subfloat[Epistemic]{\begin{centering}
\includegraphics[scale=0.45]{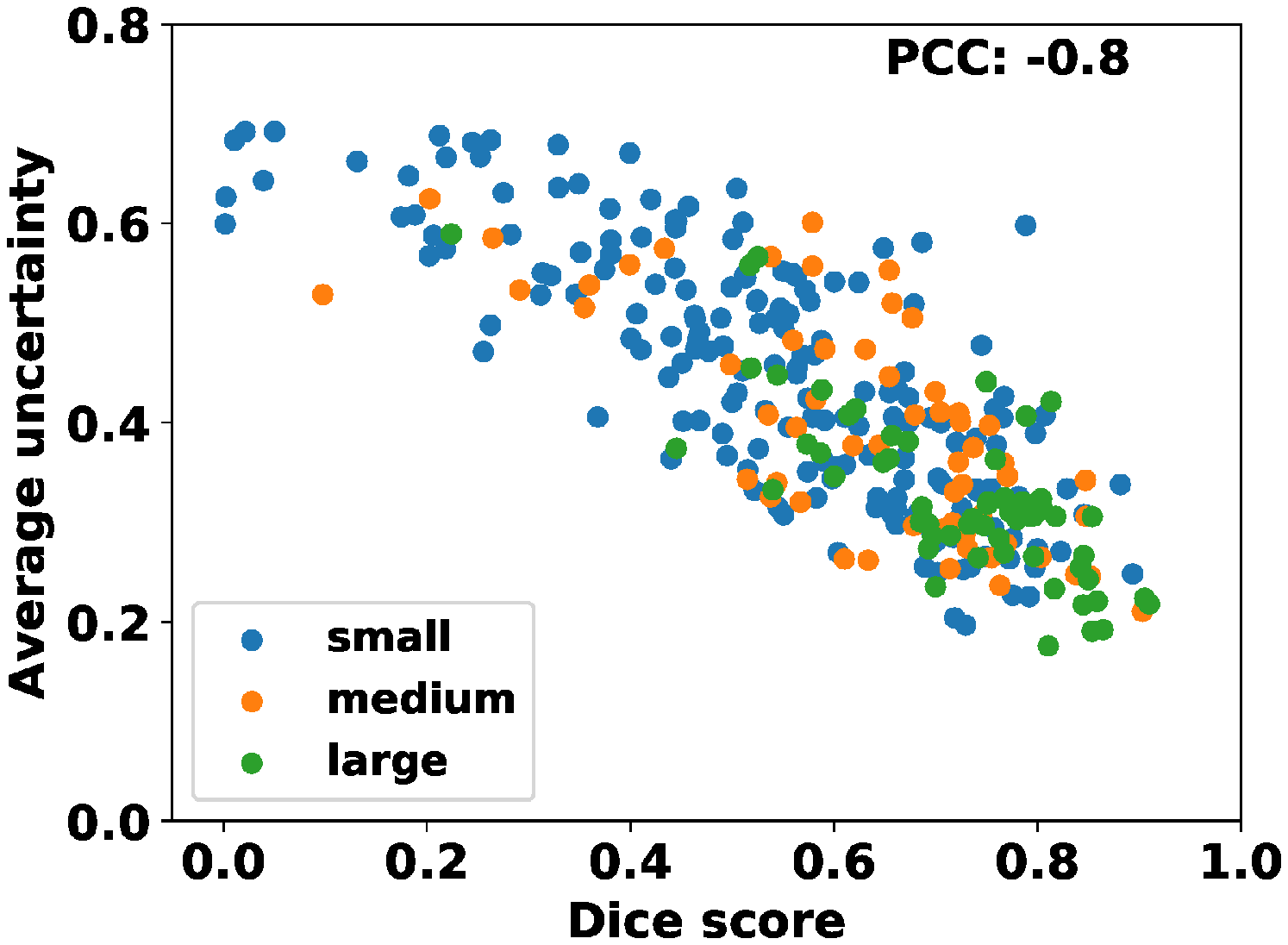} 
\par\end{centering}
\centering{} 

}\subfloat[Aleatoric]{\begin{centering}
\includegraphics[scale=0.45]{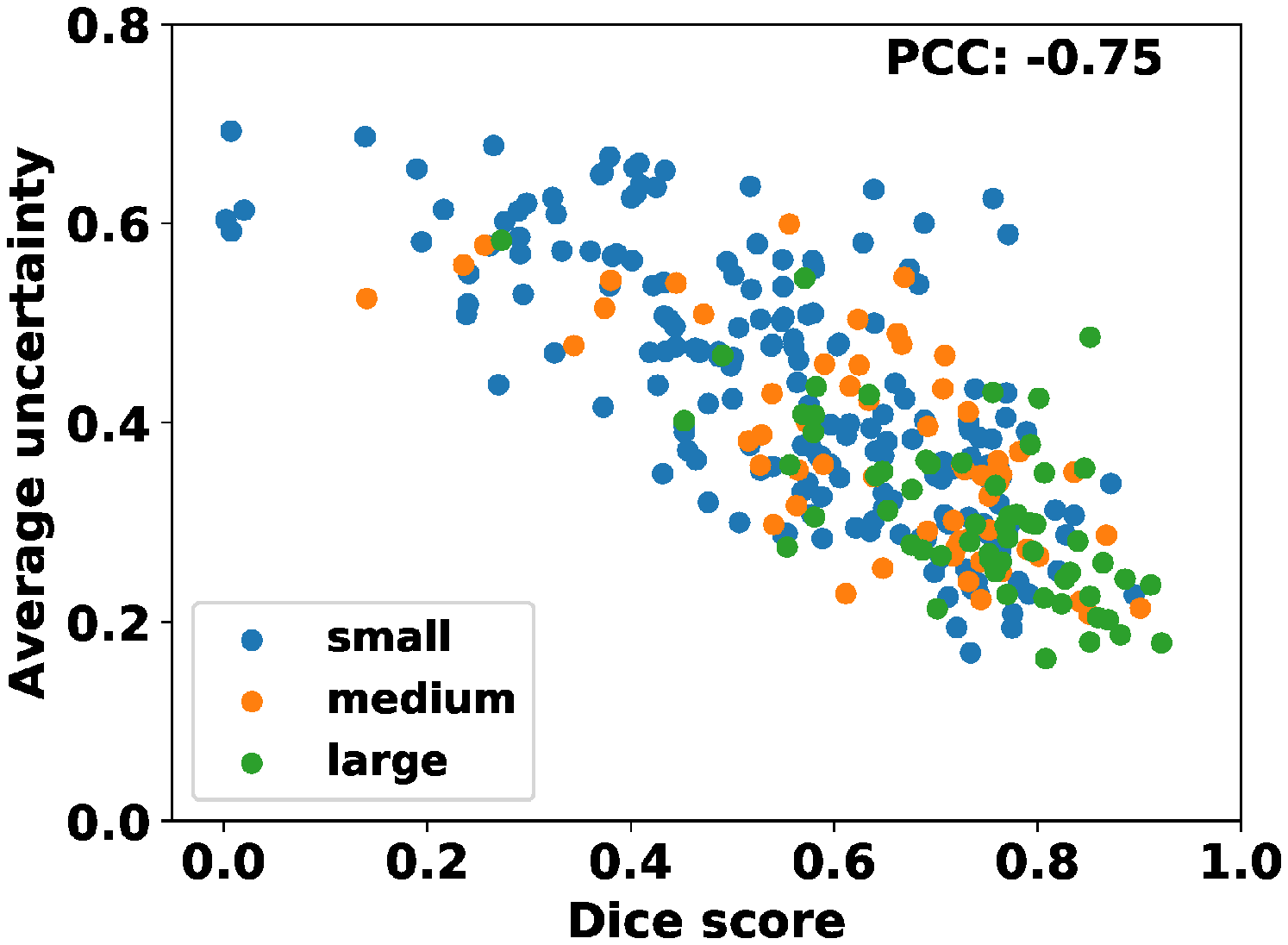} 
\par\end{centering}
\centering{} 

}
\par\end{centering}
\caption{Correlation between the average drusen segmentation uncertainty and
dice score for large, medium and small-sized drusen (marked in different
colours) on test dataset. PCC stands for Pearson correlation coefficient.\label{fig:Correlation-between-segmentation}}
\end{figure*}

In Table \ref{tab:Model-performance-on}, we report the generalization
performance of the following instances of a U-Net model: a, with no
uncertainty estimation, b, with epistemic uncertainty, and c, with
aleatoric uncertainty. The evaluation metrics include dice, precision
and recall scores. All the methods achieve similar performance. While
the models perform reasonably well on images containing large drusen,
they struggle to segment medium and small-sized drusen. These results
confirm the difficulty of the drusen segmentation task.

We now devise an evaluation strategy by utilizing the fact that higher
uncertainty can imply less confidence in model predictions. We thus
evaluate the model only on certain regions of the image by excluding
the pixels where the segmentation uncertainty is larger than a certain
threshold. We record these results in the last two rows of the Table
\ref{tab:Model-performance-on}. This evaluation strategy reveals
a significant improvement in the model performance which is reflected
across different metrics in Table \ref{tab:Model-performance-on}.
On average, this evaluation method has excluded 2-3\% of pixels that
reported higher uncertainty. Such an evaluation strategy can essentially
help us avoid evaluating the model on noisy regions (either poor-quality
data or annotation inconsistencies) of the image.

We visualize the drusen segmentation and pixel-wise epistemic and
aleatoric uncertainty in Figure \ref{fig:Visualization-of-epistemic}
along with the input test images and ground truth labels. As one would
expect the segmentation uncertainties are higher along the borders
of the drusen. Figure \ref{fig:Visualization-of-epistemic} shows
that segmentation uncertainty is relatively high for the small-sized
drusen compared to larger ones. Figure \ref{fig:Visualization-of-epistemic}
also reveals that the segmentation uncertainty can detect erroneous
segmentation by producing higher uncertainty in those regions. Moreover,
epistemic and aleatoric uncertainty differs in some regions of the
image implying they represent different aspects of the model and data.
These results outline the utility of uncertainty quantification methods
in identifying segmentation inconsistencies at test-time.

We now analyze the correlation between average segmentation uncertainty
and dice scores. We compute the Pearson correlation coefficients between
segmentation uncertainty and dice score and visualize the results
in Figure \ref{fig:Correlation-between-segmentation}. The results
show that the segmentation uncertainty exhibits strong negative linear
correlation with dice score where lower uncertainties are associated
with images having higher scores and vice versa. This analysis also
explains the poor performance of the model on medium and small-sized
drusen where both the epistemic and aleatoric uncertainty exhibits
strong negative correlation with dice scores. Figure \ref{fig:Correlation-between-segmentation}
further illustrates that the uncertainty, in general, can help explain
incorrect drusen segmentation irrespective of their size. From the
correlation plots and coefficients, epistemic uncertainty seems to
exhibit stronger negative correlation with accuracy than aleatoric.
In short, these results demonstrate that the uncertainty measures
can be helpful in providing more insights into model predictions.

\section{Conclusion\label{sec:conc}}

This paper presents the results from an initial attempt at tackling
the challenging task of segmenting the drusen in OCT images for the
early detection of AMD. We investigate the usefulness of quantifying
both the epistemic and aleatoric uncertainty in our segmentation task.
We evaluate the generalization performance of the segmentation model
in detecting large, medium and small-sized drusen. Utilizing the pixel-wise
segmentation uncertainty, we show its significance in developing a
robust evaluation framework. The visualization of the drusen segmentation
and associated uncertainty measures confirm the utility of quantifying
uncertainty in inspecting segmentation results at test-time. Our analysis
of the relationship between segmentation uncertainty and accuracy
reveals a strong negative correlation implying the utility of uncertainty
measures in identifying and explaining incorrect segmentation.%here is a scope for further
% investigation into leveraging these uncertainties to 
 %, and leverage that to build a robust segmentation model
% for drusen of various size.

On a broader note, clinical machine learning community can benefit
from the quantification of uncertainty in building robust and trustworthy
models. In particular, it can offer robust ways of evaluating clinical
machine learning models and provide additional insights to clinical
decision makers as they process the results generated by AI-powered
system. It also provides a utility to develop cost-effective training
strategies like active learning that can consume the uncertainty measures
to selectively sample the training data. %\vspace{-0.15cm}
%we believe these are the initial steps towards building a
%system for the early detection and differential diagnosis of AMD.
%\newpage 
  %\vspace{3cm}
\bibliographystyle{aaai21}
\bibliography{reference}
 
\end{document}